\documentclass[runningheads]{llncs}
\usepackage{graphicx}
\usepackage[colorlinks=true,linkcolor=blue,urlcolor=blue,citecolor=blue]{hyperref}
\renewcommand\UrlFont{\color{blue}\rmfamily}

\usepackage[ruled,vlined,linesnumbered]{algorithm2e}
\usepackage{color}
\usepackage{amssymb}
\usepackage{amsmath}
\usepackage{booktabs}
\usepackage{multirow}
\usepackage{multicol}
\usepackage{subfig}
\usepackage{pgfplots}
\pgfplotsset{width=6cm,compat=1.14}
\usepackage{float}

\newcommand{\modelname}{LGLMF\xspace}

\begin{document}
\title{LGLMF: Local Geographical based Logistic Matrix Factorization Model for POI Recommendation}
\titlerunning{Local Geographical Matrix Factorization for POI Recommendation}
%
\author{
Hossein A. Rahmani\inst{1} \and
Mohammad Aliannejadi\inst{2} \and
Sajad Ahmadian\inst{3} \and
\\
Mitra Baratchi\inst{4} \and
Mohsen Afsharchi\inst{1} \and
Fabio Crestani\inst{2}
}
\authorrunning{H. A. Rahmani et al.}
%
\institute{
University of Zanjan, Zanjan, Iran \\
\email{\{srahmani,afsharchi\}@znu.ac.ir} \and
Universit\`a della Svizzera Italiana, Lugano, Switzerland \\
\email{\{mohammad.alian.nejadi,fabio.crestani\}@usi.ch} \and
Kermanshah University of Technology, Kermanshah, Iran \\
\email{s.ahmadian@znu.ac.ir} \and
Leiden University, Leiden, The Netherlands \\
\email{m.baratchi@liacs.leidenuniv.nl}
}
\maketitle              
\begin{abstract}
With the rapid growth of Location-Based Social Networks, personalized Points of Interest (POIs) recommendation has become a critical task to help users explore their surroundings. Due to the scarcity of check-in data, the availability of geographical information offers an opportunity to improve the accuracy of POI recommendation. Moreover, matrix factorization methods provide effective models which can be used in POI recommendation. However, there are two main challenges which should be addressed to improve the performance of POI recommendation methods. First, leveraging geographical information to capture both the user's personal, geographic profile and a location's geographic popularity. Second, incorporating the geographical model into the matrix factorization approaches. To address these problems, a POI recommendation method is proposed in this paper based on a Local Geographical Model, which considers both users' and locations' points of view. To this end, an effective geographical model is proposed by considering the user's main region of activity and the relevance of each location within that region. Then, the proposed local geographical model is fused into the Logistic Matrix Factorization to improve the accuracy of POI recommendation. Experimental results on two well-known datasets demonstrate that the proposed approach outperforms other state-of-the-art POI recommendation methods.
\end{abstract}
\section{Introduction}
With the spread of smartphones and other mobile devices, Location-Based Social Networks (LBSNs) have become very popular. Therefore, LBSNs are receiving considerable attention not only for users but also from academia and industry. In LBSNs, users can share their experiences via check-ins to Points of Interests (POIs) about locations\footnote{In this paper, we use the terms location and POI interchangeably with locations.} where they have visited, such as restaurants, tourists spots and stores. Generally, the main task of POI recommendation is to recommend new and interesting POIs to users leading to improve the users' experience.

Much research has addressed POI recommendation by employing traditional recommendation methods such as Matrix Factorization (MF). MF obtains users' and POIs' latent factors based on the user-location frequency matrix, which shows the number of check-ins of users to POIs \cite{cheng2012fused,johnson2014logistic}. Due to the lack of check-in data, the MF-based POI recommendation methods suffer from data sparsity problem \cite{ahmadian2019sparsity,ahmadian2018adaptive,li2015rank,ye2011exploiting}. This problem refers to the sparsity of the user-POI matrix because the users mainly provide a few check-ins in their history. To address this problem and improve the accuracy of POI recommendation, other contextual information such as geographical, temporal, and categorical have been incorporated in the recommendation process \cite{cheng2012fused,liu2017experimental,aliannejadi2017venue}. The analysis of users' behavior indicates that geographical information has a higher impact on users' preference than other contextual information \cite{stepan2016incorporating,zhang2015geosoca,aliannejadi2018collaborative}. As a consequence, several POI recommendation methods have been proposed considering the geographical context~\cite{cheng2012fused,guo2018location,guo2019location,zhang2013igslr}. However, the past work has considered geographical context only from the user's point of view, that is, the geographical influence is based on the distance between the user's location and POIs \cite{cheng2012fused,guo2018location,guo2019location,zhang2013igslr}.

In this paper, we propose a new POI recommendation method which takes the geographical context from both users' and locations' perspectives to provide an effective geographical model. To this end, the user's high activity region is considered as the user's point of view for the proposed model. Furthermore, to model the location's point of view, we assume that the more check-ins around an unvisited POI, the less relevant this POI should be for recommendation. Moreover, the proposed geographical model is fused as a novel matrix factorization framework to improve the accuracy of recommendations. For this purpose, the proposed model is added to the logistic matrix factorization approach to propose a novel geographical-based POI recommendation. Experimental results on two real-world datasets demonstrate that considering the proposed geographical model into the matrix factorization approach achieves higher performance compared to other POI recommendation methods. This paper's contributions can be summarized as follows:
\begin{itemize}
    \item A new geographical model is proposed that considers both of the users' and the locations' perspectives.
    \item A novel POI recommendation approach is proposed by fusing the geographical model into the logistic matrix factorization approach. 
    \item The sparsity problem is addressed in the proposed method by modeling the geographical influence as an important contextual information.
    \item Several experiments are conducted on two well-known datasets demonstrating the improvement of the proposed method in the accuracy of POI recommendation compared to other state-of-the-art approaches. 
\end{itemize}

The remainder of the paper is organized as follows. The next section introduces relevant prior work on POI recommendation and contextual information. Section \ref{sec:method} gives a detailed introduction about modeling geographical information based on users' and locations' points of view. We conduct experiments and the results are presented in Sect. \ref{sec:experiments}. Finally, we conclude the paper in Sec. \ref{sec:conclusion}.

\section{Related Work}
\label{sec:relatedwork}
POI recommendation approaches mainly applies two types of techniques, including memory-based and model-based collaborative filtering into recommendation process. Memory-based approaches use users' check-in data in POI recommendation to predict users' preferences. One of the most important problems of these methods is data sparsity when a large number of elements in check-in data are empty (i.e. they do not provide any information) \cite{li2015rank,liu2017experimental}. On the other hand, several previous research is based on model-based approaches such as matrix factorization to improve the accuracy and scalability of POI recommendation \cite{griesner2015poi}. However, since there are a lot of available locations and a single user can visit only a few of them, CF-based approaches often suffer from data sparsity. As a consequence, the user-item matrix of CF becomes very sparse, leading to poor performance in cases where there is no signiﬁcant association between users and items. Many studies have tried to address the data sparsity problem of CF by incorporating additional information into the model \cite{cheng2012fused,ye2011exploiting}.

Furthermore, check-in data often include several significant contextual information such as geographical, temporal, categorical, and textual and matrix factorization methods attempt to consider such information to improve the quality of recommendations \cite{aliannejadi2018personalized,aliannejadi2016user,liu2013personalized,rahmani2019category}. Rahmani et al. \cite{rahmani2019category} propose a POI embedding model, CATAPE, that take into account the characteristics of POI by POI categories. CATAPE consists of two modules, Check-in module and Category module, to incorporate the user's sequence behaviour and location's properties.

It has been shown that geographical context is an important factor which considers location changes of users between POIs \cite{liu2017experimental}. Past research has addressed how to model this type of information in recommendation process \cite{cheng2012fused,li2015rank,ye2011exploiting}. Ye et al.~\cite{ye2011exploiting}, showed that POIs visited by the users follow the Power-law Distribution (PD) when the geographical information is considered. Moreover, they proposed a memory-based CF method which suffers from a scalability problem about large-scale data. Cheng et al.~\cite{cheng2012fused} proved that users' check-ins revolve around multiple centers which are captured using a Multi-center Gaussian Model (MGM). Li et al.~\cite{li2015rank} considered the task of POI recommendation as to the problem of pair-wise ranking, where they exploited the geographical information using an extra factor matrix. Zhang et al.~\cite{zhang2013igslr} proposed a method that considered the geographical influence on each user separately. To this end, a model is proposed based on Kernel Density Estimation (KDE) of the distance distributions between POIs checked-in by each user. Yuan et al. \cite{yuan2016joint} addressed the data sparsity problem assuming that users tend to rank higher the POIs that are geographically closer to the one that they have already visited.

More recently, Aliannejadi et al.~\cite{aliannejadi2019joint} propose a two-phase collaborative ranking algorithm for POI recommendation. They push POIs with single or multiple check-ins at the top of the recommendation list, taking into account the geographical influence of POIs in the same neighborhood. They show that both visited, and unvisited POIs in the learning alleviates the sparsity problem. Guo et al.~\cite{guo2019location} proposed a Location neighborhood-aware Weighted Matrix Factorization (L-WMF) model that incorporate the geographical relationships among POIs to exploit the geographical characteristics from a location perspective.

The previous approaches mainly explored the geographical information from a user's perspective. In comparison to the other POI recommendation models, the method proposed in this paper is based on combining the users' and locations' points of view into a better geographical information model. To this end, the distance between users and POIs (from the user's point of view) and the check-in frequency on neighboring POIs (from the location's point of view) are used in the proposed model. Moreover, we address data sparsity taking into account the influence of POIs' neighbors in the recommendation strategy of our model.

\section{Proposed Method}
\label{sec:method}
This section presents the proposed POI recommendation method called Local Geographical based Logistic Matrix Factorization (LGLMF). LGLMF consists of two main steps. In the first step, a Local Geographical Model (LGM) is proposed based on both users' and locations' points of view. Then, in the second step, the LGM is fused into a Logistic Matrix Factorization (LMF) approach. The fused matrix factorization model is used to predict the users' preferences. The details of the main steps of the proposed method are provided in the following subsections.
\subsection{Local Geographical Model}
In this section, the proposed local geographical model is introduced. Let $U=\{u_1, u_2, u_3, ..., u_m\}$ be the set of users and $P=\{p_1, p_2, p_3, ..., p_n\}$ be the set of POIs in a typical LBSN. Then, let $C\in{\mathbb{R}^{m\times{n}}}$ be a user-POI check-in frequency matrix with $m$ users and $n$ POIs. The value $c_{up}\in{C}$ show the check-in frequency of user $u$ to the POI $p$. Then, the task of personalized top-N POI recommendation problem is formally defined as follows:

\textsc{Definition 1 (Top-N POI Recommendation).} \textit{Given a user-POI check-in frequency matrix $C$ and a set of POIs $P^u\subseteq{P}$ that have been visited by the user $u$, identify $X=\{p_1, p_2, ..., p_N\}$, a set of POIs ordered based on the probability of a user's visit in the future such that $|X|\leq{N}$ and $X\cap{P^u}=\varnothing$.}

The proposed geographical model captures the geographical influence of both users' and locations' points of view. From the user's point of view, the geographical information can be modeled by considering the user's activity region. On the other hand, from the location's point of view, the geographical information can be modeled as the number of check-ins on the neighbors of a selected POI. Therefore, it can be indicated how agreeable a location is relative to its neighbors. The pseudo code of the proposed LGM is presented in Algorithm \ref{alg:LBP}. The algorithm is composed of three inner loops to model the geographical information, the first two loops model the user's region (lines $2-5$) and the third loop calculates the probability of a user preferring a POI within a neighborhood, considering the visits to its neighboring POIs (lines $6-10$). For modeling the user's region, we need to find each user's high activity location (in the real world, this could be the user's residence region). To this end, the user's most frequently checked-in POI is taken to infer his/her high activity location (line $1$). Then, we scan the list of unvisited POIs to find those that fall in the same region (in-region) for that user, that is laying within $\alpha$ kilometers from the user's high activity location (the user's perspective) (line $5$). Moreover, based on in-region POIs for each user, we consider the impact of checked-in neighboring POIs, whose distance is less than $\gamma$ meters from the unvisited POIs (the location's perspective) (line $7-10$). The POI locality is defined as follows:

\textsc{Definition 2 (POI Locality).} \textit{Given a set of POIs $P=\{p_1, p_2, ..., p_n\}$, each $p_i$ $(p_i\in{P})$ has a POI Locality with respect to user $u$ denoted as $p_i^u$ (Eq. \ref{eq:POIlocality}), which is the user $u's$ preference on POI $p_i$ relative to its neighbors.}
\begin{equation}
p_i^u=1 - \frac{L_p^u}{|P^u|}
\label{eq:POIlocality}
\end{equation}
Here, $L_p^u$ denotes the number of neighbors of $p_i$ visited by user $u$. Also, $|P^u|$ is the cardinality of the set of POIs that user $u$ has visited.

\begin{algorithm}
	\DontPrintSemicolon
	\SetAlgoLined
	\SetKwInOut{Input}{Input}\SetKwInOut{Output}{Output}
	\Input{\textit{U, P, $\alpha$, $\gamma$}}
	\Output{user-POI preference matrix {$\hat{M}$}}
	\tcc{$\hat{M}$ is a $U\times{P}$ matrix, and all elements are intialized by $0$.}
	HAL $\leftarrow$ Find each user's high activity location \;
	\tcc{User's most frequently checked-in POI is taken as HAL}
	\ForEach{$u\in{U}$}{    
		\ForEach{$p\in{P}$}{
			\If{$p\notin{P^u}$}{
				\If{distance($p$, HAL\textsubscript{u}) $< \alpha$}{
					\ForEach{$p^u\in{P^u}$}{					
						\If{distance($p$, $p^u$) $< \gamma$}{
							$L_p^u\leftarrow L_p^u + 1$    \;
						}
					$\hat{M}[u,p] \leftarrow 1 - \frac{L_p^u}{|P^u|}$  (Eq. \ref{eq:POIlocality})\;
					}					
				}
			}
		}
	}
	\Return{$\hat{M}$}
	\caption{The Proposed Local Geographical Model}
	\label{alg:LBP}
\end{algorithm}

\subsection{Constructing the Matrix Factorization Model}
Traditional recommender systems (RSs) mainly rely on explicit feedback data as input. The explicit feedback data includes the preferences of users about the existing items. For example, in Netflix, users can express their preferences about movies by providing star ratings. Frequently, explicit feedback data does not exist in LBSNs. Therefore, check-in data can be considered as implicit feedback data for RSs, forming a different recommendation problem \cite{johnson2014logistic}. 

C.C. Johnson in \cite{johnson2014logistic} previously proposed a Logistic Matrix Factorization (LMF) model, achieving a significant result with the implicit feedback dataset of Spotify music. The LMF takes a probabilistic approach that models the probability of a user's preference on an item by a logistic function. However, LMF fails to consider contextual information into the recommendation process. In this section, a novel matrix factorization method is proposed based on LMF by considering the proposed LGM as additional contextual information. Generally, the aim of MF-based recommendation is to find two low-rank matrices including the user-factors matrix $V\in{\mathbb{R}^{m\times{k}}}$ and item-factors matrix $L\in{\mathbb{R}^{m\times{k}}}$ where $k$ is the number of latent factors such that the inner product of these two matrices approximate matrix $\hat{C}$, i.e. $\hat{C}=V\times{L^{T}}$. Each row of $V$ ($v_u\in{V}$) represents a user's vector of user's behaviour and each row of $L$ ($l_p\in{L}$) represents the act of item's properties.

Suppose $e_{u,p}$ denotes the number of check-ins by user $u$ on POI $p$ (user $u$ prefers POI $p$), and the parameters $V$ and $L$ are two latent factors for users and POIs, respectively. Also, consider $\beta_u$ and $\beta_p$ as user bias and POI bias. The probability $P(e_{up}|v_u,l_p,\beta_u,\beta_p)$ is defined to represent the preference of user $u$ on POI $p$ as follows:

\begin{equation}
P(e_{up}|v_u,l_p,\beta_u,\beta_p)= \frac{exp(v_ul_p^T + \beta_u + \beta_p)}{1+exp(v_ul_p^T + \beta_u + \beta_p)}
\label{eq:LMF}
\end{equation}
Moreover, the parameters $V$, $L$, and $\beta$ can be learned by solving the following optimization problems:
\begin{equation}
arg \ max_{V, L, \beta} \ logP(V, L, \beta| C)
\end{equation}
where $logP(V, L, \beta| C)$ is defined as follows:
\begin{equation}
\begin{aligned}
\sum_{u,p}\alpha{c_{up}}(v_ul^{T}_p + \beta_{u} + \beta_{p})-(1+\alpha{c_{up}}) log(1+ \\ exp(v_ul^{T}_p + \beta_{u} + \beta_{p}))-\frac{\lambda}{2}||v_u||^2-\frac{\lambda}{2}||l_p||^2
\end{aligned}
\end{equation}

Finally, we fuse the proposed LGM (i.e., Alg. \ref{alg:LBP}) into the matrix factorization method. Therefore, the probability of a user $u$ visiting a POI $p$ can be calculated as follows:

\begin{equation}
\label{eq:5}
Preference_{up}=P(e_{up}|v_u,l_p,\beta_u,\beta_p) \times{\hat{M}(u,p)}
\end{equation}

where $P(e_{up}|v_u,l_p,\beta_u,\beta_p)$ is calculated by Eq. \ref{eq:LMF} and $\hat{M}(u,p)$ is calculated by Alg. \ref{alg:LBP}. A list of POI recommendations can be provided for each user by using the proposed probability function (i.e. Eq. \ref{eq:5}). It should be noted that, differently from the LMF method, the proposed LGLMF model considers the contextual information into the recommendation process by fusing the proposed LGM.

\section{Experiments}
\label{sec:experiments}
In this section, several experiments are conducted to compare the performance of LGLMF with the other POI recommendation methods. The details of the experiments are discussed in the following subsections.

\subsection{Experimental Settings}
\subsubsection{Datasets.}
We evaluated the algorithms using two real-world check-in datasets\footnote{\UrlFont{http://spatialkeyword.sce.ntu.edu.sg/eval-vldb17/}} collected from Gowalla and Foursquare \cite{liu2017experimental}.
Gowalla includes check-ins from February 2009 to October 2010, while Foursquare includes check-in data from April 2012 to September 2013. Each check-in contains a user, a POI (latitude and longitude), and the check-in timestamp. Users with less than 15 check-in POIs and POIs with less than ten visitors have been removed from Gowalla. On the other hand, users with less than ten check-in POIs and also POIs with less than ten visitors have been removed from Foursquare. The statistical details of the datasets are presented in Table \ref{tbl:datasets}.

\bgroup
\def\arraystretch{1.5}
\begin{table}[h]
\centering
\caption{Statistics of the evaluation datasets}
\label{tbl:datasets}
\begin{tabular}{l|c|c|c|c}
\hline
\textbf{Datasets} & \textbf{\#Users  } & \textbf{\#POIs  } & \textbf{\#Check-ins  } & \textbf{Sparsity  } \\ \hline
Gowalla           & 5,628             & 31,803           & 620,683               & 99.78\%           \\ \hline
Foursquare        & 7,642             & 28,483           & 512,523               & 99.87\%           \\ \hline
\end{tabular}
\end{table}

\subsubsection{Evaluation Metrics.}
Three ranking-based evaluation metrics including Pre@$N$ (Precision at $N$), Rec@$N$ (Recall at $N$), and nDCG@$N$ with $N \in \{10, 20\}$ are used to evaluate the performance of the recommendation methods.
$Pre@N$ refers to the ratio of recovered POIs to the N recommended POIs and $rec@N$ refers to the ratio of recovered POIs to the number of POIs predicted by the recommendation model. Moreover, $nDCG@N$ is a measure to indicate the ranking quality of the recommendation models.
We partition each dataset into training data, validation data, and test data. For each user, we use the earliest $70\%$ check-ins as the training data, the most recent $20\%$ check-ins as the test data and the remaining $10\%$ as the validation data. We determine the statistically significant differences using the two-tailed paired t-test at a $95\%$ confidence interval ($p < 0.05$).

\subsubsection{Comparison Methods.}
The proposed \modelname model is compared with the POI recommendation approaches that consider geographical influence in the recommendation process. Moreover, the POI recommendation models which are based on the geographical information from the locations' points of view are considered in the experiments. The details of the compared methods are listed as follows:

\begin{itemize}
    \item LMF \cite{johnson2014logistic}: A Logistic Matrix Factorization method that incorporates a logistic function.
    \item PFMMGM \cite{cheng2012fused}: A method based on the observation that user's check-in around several centers, that applies Multi-center Gaussian Model (MGM) to study user's behavior.
    \item LRT \cite{gao2013exploring}: A model that incorporates temporal information in a latent ranking model and learns the user's preferences to locations at each time slot.
    \item PFMPD: A method using the Power-law Distribution \cite{ye2011exploiting} that model people tend to visit nearby POIs. We integrate this model with the Probabilistic Factor Model (PFM).
    \item LMFT \cite{stepan2016incorporating}: A method that considers a user's recent activities as more important than their past activities and multiple visits to a location, as indicates of a stronger preference for that location.
    \item iGLSR\footnote{We evaluate iGLSR only on Gowalla as we do not have access to the social data of the Foursquare dataset.
    } \cite{zhang2013igslr}: A method that personalizes social and geographical influence on location recommendation using a Kernel Density Estimation (KDE) approach.
    \item L-WMF \cite{guo2019location}: A location neighborhood-aware weighted probabilistic matrix factorization (L-WMF) model that incorporates the geographical relationships among POIs into the WMF as regularization to exploit the geographical characteristics from a location perspective.
    \item \modelname\footnote{\UrlFont{\href{https://github.com/rahmanidashti/LGLMF}{https://github.com/rahmanidashti/LGLMF}}}: Our proposed method that fused LMF with the proposed local geographical model.
\end{itemize}

\subsubsection{Parameter Settings.}
For the baseline methods, the parameters are initialized as reported in the corresponding papers. We set the latent factors parameter as $k = 30$ for LMF, PFM, L-WMF. For PFGMGM, we set the distance threshold $d$ to $15$ and the frequency control parameter $\alpha$ to $0.2$ based on the original paper. For LRT, we set $T$ as temporal states to $24$ and the $\alpha$ and $\beta$ as regularization parameters to $2.0$. We tune the \modelname parameters based on the validation data. We find the optimal values for the parameters using the validation data and use them in the test data.

\subsection{Performance Comparison}
Table \ref{tab:pre_yelp} shows the results of experiments based on the Gowalla and Foursquare datasets. As you can see from these results, \modelname obtains higher accuracy than the other POI recommendation methods based on both of the datasets. Therefore, it can be concluded that incorporating the contextual information into the matrix factorization leads to improve the quality of POI recommendation. In comparison to PFMPD, PFMMGM, and iGLSR as the three basic approaches in modeling the geographical influence, \modelname achieves better results for both datasets based on all metrics. Among these baselines, iGLSR performs better. This is because iGLSR models geographical influence based on each user's behavior.
\begin{table*}[t]
  \caption{Performance comparison with baselines in terms of Pre@$k$, Rec@$k$, and nDCG@$k$ for $k \in \{10,20\}$ on Gowalla and Foursquare. The superscript $\dagger$ denotes significant improvements compared to baselines ($p < 0.05$).}
  \label{tab:pre_yelp}
  \begin{tabular}{llllccccc}
    \toprule
    \multirow{2}{*}{\textbf{Dataset}} & \multirow{2}{*}{\textbf{Method}} & \multicolumn{6}{c}{\textbf{Metrics}} \\
    \cmidrule{3-8}
    & & Pre@10 & Pre@20 & Rec@10 & Rec@20 & nDCG@10 & nDCG@20 \\
    \midrule
    \multirow{8}{*}{\textbf{Gowalla}} & LMF & 0.0328 & 0.0272 & 0.0325 & 0.0534 & 0.0167 & 0.0159 \\
    & PFMMGM & 0.0240 & 0.0207 & 0.0258 & 0.0442 & 0.0140 & 0.01440 \\
    & LRT & 0.0249 & 0.0182 & 0.0220 & 0.0321 & 0.0105 & 0.0093 \\
    & PFMPD & 0.0217 & 0.0184 & 0.0223 & 0.0373 & 0.0099 & 0.0101 \\
    & LMFT & 0.0315 & 0.0269 & 0.0303 & 0.0515 & 0.0157 & 0.0150 \\
    & iGLSR & 0.0297 & 0.0242 & 0.0283 & 0.0441 & 0.0153 & 0.0145 \\
    & L-WMF & 0.0341 & 0.0296 & 0.0351 & 0.0582 & 0.0183 & 0.0178 \\
    & \textbf{\modelname} & \textbf{0.0373}$^{\dagger}$ & \textbf{0.0317}$^{\dagger}$ & \textbf{0.0383}$^{\dagger}$ & \textbf{0.0629}$^{\dagger}$ & \textbf{0.0212}$^{\dagger}$ & \textbf{0.0208}$^{\dagger}$ \\
  \midrule
    \multirow{8}{*}{\textbf{Foursquare}} & LMF & 0.0228 & 0.0189 & 0.0342 & 0.0565 & 0.0136 & 0.0148 \\
    & PFMMGM & 0.0170 & 0.0150 & 0.0283 & 0.0505 & 0.0109 & 0.0126 \\
    & LRT & 0.0199 & 0.0155 & 0.0265 & 0.0425 & 0.0117 & 0.0124 \\
    & PFMPD & 0.0214 & 0.0155 & 0.0290 & 0.0426 & 0.0124 & 0.0128 \\
    & LMFT & 0.0241 & 0.0194 & 0.0359 & 0.0568 & 0.0150 & 0.0161 \\
    & L-WMF & 0.0248 & 0.0197 & 0.0387 & 0.0591 & 0.0162 & 0.0174 \\
    & \textbf{\modelname} & \textbf{0.0266}$^{}$ & \textbf{0.0213}$^{\dagger}$ & \textbf{0.0424}$^{\dagger}$ & \textbf{0.0678}$^{\dagger}$ & \textbf{0.0175}$^{}$ & \textbf{0.0192}$^{\dagger}$ \\
  \bottomrule
\end{tabular}
\end{table*}
It should be noted that previous models only consider the user's point of view for the geographical influence. Also, our proposed model outperforms the L-WMF that is state-of-the-art model that uses location's prospective geographical information. Compared to the state-of-the-art method, L-WMF, the improvements in terms Rec@20 and nDCG@20 on Gowalla are $8\%$ and $15\%$, respectively. Therefore, the results confirm the effectiveness of our LGM model, which considers both of the users' and the locations' points of view in modeling the geographical influence.

\subsubsection{Impact of number of visited POIs.}
Table \ref{tbl:sparsity} shows Rec@$20$ and nDCG@$20$ of all models based on different percentages of POIs that each user has visited in the training data. These results indicate that \modelname achieves the highest accuracy in comparison to the other recommendation models for different number of POIs. Therefore, it is shown that \modelname can address the sparsity problem where the training data is not enough to provide reliable recommendations. Also, we observe a more robust behavior of \modelname compared to the baselines. Thus, the proposed local geographical model enables \modelname to deal with noise and data sparsity effectively. This is clearer when \modelname outperforms other methods with a larger margin in terms of Rec@$20$.

\begingroup
\setlength{\tabcolsep}{2pt}
\begin{table*}[t]
  \caption{Effect on Rec@20 and nDCG@20 of different number of POIs that users visited as training data on Gowalla and Foursquare. The superscript $\dagger$ denotes significant improvements compared to all baselines ($p < 0.05$).}
  \label{tbl:sparsity}
  \begin{tabular}{llccccccc}
    \toprule
    \multirow{2}{*}{\textbf{Dataset}} & \multirow{2}{*}{\textbf{Method}} & \multicolumn{3}{c}{\textbf{Rec@20}} &
    & 
    \multicolumn{3}{c}{\textbf{nDCG@20}} \\
    \cmidrule{3-5} \cmidrule{7-9}
    & & $40\%$ & $60\%$ & $80\%$ && $40\%$ & $60\%$ & $80\%$ \\
    \midrule
    \multirow{8}{*}{\textbf{Gowalla}} & LMF & 0.0205 & 0.0333 & 0.0455 && 0.0058 & 0.0095 & 0.0129 \\
    & PFMMGM & 0.0350 & 0.0376 & 0.0414 && 0.0094 & 0.0107 & 0.0120 \\
    & LRT & 0.0014 & 0.0318 & 0.0300 && 0.0003 & 0.0098 & 0.0083 \\
    & PFMPD & 0.0235 & 0.0323 & 0.0346 && 0.0064 & 0.0093 & 0.0088 \\
    & LMFT & 0.0205 & 0.0321 & 0.0426 && 0.0057 & 0.0088 & 0.0119 \\
    & iGLSR & 0.0317 & 0.0357 & 0.0405 && 0.0097 & 0.0105 & 0.0128 \\
    & L-WMF & 0.0382 & 0.0435 & 0.0471 && 0.0099 & 0.0110 & 0.0142 \\
    & \textbf{\modelname} & \textbf{0.0479}$^{\dagger}$ & \textbf{0.0533}$^{\dagger}$ & \textbf{0.0580}$^{\dagger}$ && \textbf{0.0137}$^{\dagger}$ & \textbf{0.0158}$^{\dagger}$ & \textbf{0.0187}$^{\dagger}$ \\
  \midrule
    \multirow{8}{*}{\textbf{Foursquare}} & LMF & 0.0180 & 0.0266 & 0.0442 && 0.0048 & 0.0073 & 0.0122 \\
    & PFMMGM & 0.0451 & 0.0466 & 0.0491 && 0.0097 & 0.0107 & 0.0120 \\
    & LRT & 0.0402 & 0.0434 & 0.0438 && 0.0118 & 0.0125 & 0.0125 \\
    & PFMPD & 0.0371 & 0.0389 & 0.0412 && 0.0117 & 0.0119 & 0.0124 \\
    & LMFT & 0.0180 & 0.0288 & 0.0418 && 0.0049 & 0.0077 & 0.0112 \\
    & L-WMF & 0.0473 & 0.0492 & 0.0537 && 0.0119 & 0.0131 & 0.0142 \\
    & \textbf{\modelname} & \textbf{0.0500}$^{\dagger}$ & \textbf{0.0578}$^{\dagger}$ & \textbf{0.0660}$^{\dagger}$ && \textbf{0.0126} & \textbf{0.0154}$^{\dagger}$ & \textbf{0.0189}$^{\dagger}$ \\
  \bottomrule
\end{tabular}
\end{table*}
\endgroup

\subsubsection{Impact of $\alpha$ and $\gamma$.}
Figure \ref{fig:parameters} shows the performance of \modelname based on different values of $\alpha$ and $\gamma$. In Figure \ref{fig:alpha}, the effect of different $\alpha$ values on the performance of \modelname is reported based on Rec@$20$ metric. As you can see from these results, the optimal value of $\alpha$ for both Gowalla and Foursquare datasets is $20$. Figure \ref{fig:gamma} shows the effect of different $\gamma$ values on the performance of \modelname based on Gowalla and Foursquare. It can be seen that the optimal value of $\gamma$ for both datasets is $10$. These results show that users tend to visit near locations and they make a region from their high activity location.

\begin{figure}[t]
	\centering
	\subfloat[][Effect of $\alpha$]{%
		\label{fig:alpha}%
		\begin{tikzpicture}
        \begin{axis}[
            axis lines = left,
            xlabel = $\alpha$,
            ylabel = {Rec@$20$},
            legend pos=south east,
            xmin=0, xmax=100,
            ymin=0, ymax=0.06,
            xtick={0,20,40,60,80,100},
            ytick={0,0.01,0.02,0.03,0.04,0.05,0.06}
        ]
        \addplot [
            color=red,
            mark=halfcircle
        ]
        coordinates{
        (0,0.0003)(10,0.0463)(20,0.0491)(30,0.0468)(40,0.0474)(50,0.0461)(60,0.0445)(70,0.0458)(80,0.044)(90,0.0449)(100,0.0438)
        };
        \addlegendentry{$Gowalla$}
        \addplot [
            color=blue,
            mark=square
            ]
            coordinates{
            (0,0.0002)(10,0.0372)(20,0.0397)(30,0.0381)(40,0.0371)(50,0.0368)(600.0352)(70,0.0343)(80,0.0321)(90,0.0317)(100,0.0312)
            };
        \addlegendentry{$Foursquare$}
        \end{axis}
        \end{tikzpicture}}
	\subfloat[][Effect of $\gamma$]{%
		\label{fig:gamma}%
		\begin{tikzpicture}
        \begin{axis}[
            axis lines = left,
            xlabel = $\gamma$,
            ylabel = {Rec@$20$},
            legend pos=south east,
            xmin=5, xmax=15,
            ymin=0, ymax=0.06,
            xtick={5,7,9,11,13,15},
            ytick={0,0.01,0.02,0.03,0.04,0.05,0.06}
        ]
        \addplot [
            color=red,
            mark=halfcircle
        ]
        coordinates{
        (5,0.0411)(6,0.0421)(7,0.0418)(8,0.0421)(9,0.0432)(10,0.0498)(11,0.0481)(12,0.0471)(13,0.047)(14,0.0452)(15,0.0443)
        };
        \addlegendentry{$Gowalla$}
        \addplot [
            color=blue,
            mark=square
            ]
        coordinates{
        (5,0.0312)(6,0.0341)(7,0.0346)(8,0.0347)(9,0.0365)(10,0.0399)(11,0.0374)(12,0.0356)(13,0.0343)(14,0.0332)(15,0.0314)
        };
        \addlegendentry{$Foursquare$}
        \end{axis}
        \end{tikzpicture}}
	\caption[]{Effect of different model parameters on the performance of \modelname}%
	\label{fig:parameters}%
\end{figure}
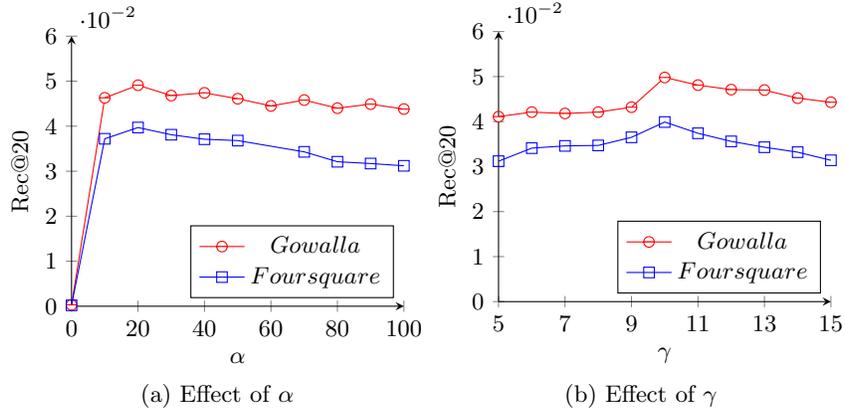

\subsubsection{Model flexibility.}
\label{sec:modelflexibility}
Our proposed geographical model can be easily fused to other POI recommendation models to improve their prediction quality. As shown in Table \ref{tbl:model_flexibility}, the proposed local geographical model is added to other models to show its impact on accuracy improvement. To this end, the proposed model is added to the LRT and PFM methods, and their performances are compared with the original models. Table \ref{tbl:model_flexibility} indicates that our geographical model has a positive impact on the performance of other POI recommendation models.

\begingroup
\setlength{\tabcolsep}{5pt}
\begin{table*}[h]
  \caption{Comparison of model flexibility with the related baselines in terms of Pre@$20$, Rec@$20$, and nDCG@$20$ on Gowalla and Foursquare. The superscript $\dagger$ denotes significant improvements compared to related baselines ($p < 0.05$).}
  \label{tbl:model_flexibility}
  \begin{tabular}{lccccccc}
    \toprule
    \multirow{2}{*}{\textbf{Method}} & \multicolumn{3}{c}{\textbf{Gowalla}} &
    & 
    \multicolumn{3}{c}{\textbf{Foursquare}} \\
    \cmidrule{2-4} \cmidrule{6-8}
    & Pre@$20$ & Rec@$20$ & nDCG@$20$ && Pre@$20$ & Rec@$20$ & nDCG@$20$ \\
    \midrule
    PFMPD & 0.0184 & 0.0373 & 0.0101 && 0.0155 & 0.0426 & 0.0128 \\
    PFMMGM & 0.0207 & 0.0442 & 0.0144 && 0.0150 & 0.0505 & 0.0126 \\
    \textbf{PFMLGM} & \textbf{0.0309}$^{\dagger}$ & \textbf{0.0588}$^{\dagger}$ & \textbf{0.0197} && \textbf{0.0198}$^{\dagger}$ & \textbf{0.0639}$^{\dagger}$ & \textbf{0.0193} \\
    \midrule
    LRT & 0.0182 & 0.0321 & 0.0093 && 0.0155 & 0.0425 & 0.0124 \\
    \textbf{LRTLGM} & \textbf{0.0330}$^{\dagger}$ & \textbf{0.0616}$^{\dagger}$ & \textbf{0.0224}$^{\dagger}$ && \textbf{0.0230}$^{\dagger}$ & \textbf{0.0717}$^{\dagger}$ & \textbf{0.0234}$^{\dagger}$
    \\
  \bottomrule
\end{tabular}
\end{table*}
\endgroup

\section{Conclusions}
\label{sec:conclusion}
In this paper, we proposed a novel Local Geographical model for POI recommendation to consider both users' and locations' point of view of geographical information. We incorporated the user's preference by Logistic Matrix Factorization and proposed a fused matrix factorization method to include the geographical information captured by our proposed Local Geographical model. Experimental results on two well-known datasets showed that the proposed method outperforms other state-of-the-art approaches through leveraging more conditions in the geographical information than competitive models. Also, in Sec. \ref{sec:modelflexibility}, we showed that LGM can be joint to other model and improve their recommendation process. Our future work will investigate how to incorporate other contextual information to our model like for example social and temporal information.

\bibliographystyle{splncs04}
\bibliography{references.bib}
\end{document}